\documentclass[conference]{IEEEtran}
\IEEEoverridecommandlockouts
\usepackage{cite}
\usepackage{amsmath,amssymb,amsfonts}
\usepackage{algorithmic}
\usepackage{graphicx}
\usepackage{textcomp}
\usepackage{blindtext}
\usepackage{romannum}
\usepackage{enumitem}
\usepackage{authblk}
\usepackage{amsmath}
\makeatletter
\def\footnoterule{\kern-3\p@
  \hrule \@width 3.5in \kern 2.6\p@} 
\makeatother

\usepackage{float}
\usepackage{xcolor}

\usepackage{stfloats}
\usepackage{mathtools, nccmath}

\usepackage{calc}
\newcommand{\RNum}[1]{\uppercase\expandafter{\romannumeral #1\relax}}
\usepackage[numbers,sort&compress]{natbib}
\def\BibTeX{{\rm B\kern-.05em{\sc i\kern-.025em b}\kern-.08em
    T\kern-.1667em\lower.7ex\hbox{E}\kern-.125emX}}
\begin{document}

\title{Using Terminal Circuit for Power System Electromagnetic Transient Simulation
\thanks{This work is supported by the State Grid Corporation technology project 5455HJ180021.}
}

\author[1, 2]{Yijing Liu}
\author[1]{Xiang Zhang}
\author[1]{Renchang Dai}
\author[1]{Guangyi Liu}
\affil[1]{Global Energy Interconnection Research Institute, San Jose, CA 95134, USA}
\affil[2]{Department of Electrical and Computer Engineering,  Texas A\&M University, College Station, TX 77843, USA}

\maketitle

\begin{abstract}
The modern power system is evolving with increasing penetration of power electronics introducing complicated electromagnetic phenomenon. Electromagnetic transient (EMT) simulation is essential to understand power system behavior under disturbance which however is one of the most sophisticated and time-consuming applications in power system. To improve the electromagnetic transient simulation efficiency while keeping the simulation accuracy, this paper proposes to model and simulate power system electromagnetic transients by very large-scale integrated circuit (VLSI) as a preliminary exploration to eventually represent power system by VLSI circuit chip avoiding numerical calculation. To proof the concept, a simple 5 bus system is modeled and simulated to verify the feasibility of the proposed approach.
\end{abstract}


\section{Introduction}
Electromagnetic Transient (EMT) analysis plays important roles in modern power systems \cite{kundur2004definition,dommel1997techniques}. 
EMT studies consider three-phase voltage and current during simulations and the time-step is in microseconds or even less to simulate power electronics switching. With the development of modern power grids, power electronics are highly penetrated in both generation system and transmission system \cite{tolbert2005power}. 
EMT simulators model power system with details introducing long simulation time. 

Many attempts have been made to improve the computation efficiency \cite{anderson1995new,sultan1998combined,van2014advanced,gao2009frequency,yang2015dynamic,plumier2014relaxation,jalili2009interfacing,zhang2012development,heffernan1981computation,su2004recent}. But it’s challenging in the fact of that the conventional numerical method is adopted to solve the differential algebraic equations representing power system electromagnetic transient by time domain simulation which consumes unaffordable time for large-scale systems. Physically, power system is a very large-scale circuit consisting resistors, inductors, reactors, and other circuit elements. Using integrated circuit chip to mimic power system with high fidelity in real time could leverage the advanced technology of integrated circuit industry in terms of modelling, simulation, and manufacturing. In work \cite{zhang2018vlsi}, a very large-scale integration (VLSI) architecture for exciters, governors, and power system stabilizers is presented and a fixed time-step approach is used for power system electromechanical transient simulation purpose. Circuit simulation results demonstrate that power system modeled by VLSI circuits can be simulated faster than real time without losing accuracy. It’s worth noting that the power grid can be essentially simplified as a linear network composed by resistors, inductors and capacitors when investigating the node voltages with the node current injections. The authors of \cite{dai2019simplify} take the advantage of the linearity and propose a method to simplify power flow calculation by utilizing terminal circuit model and PMU measurements. This approach is further extended in work \cite{zhang2021dynamic}, which proposed a new method to model power systems as an integrated circuit and simulate the system dynamics by integrated circuit simulator. In this paper, the terminal circuit method is further investigated and applied to transient simulation. The research goal of this paper is to extend the terminal circuit based steady-state and dynamic cases with appropriate power system equipment electromagnetic models, to study electromagnetic transients under normal or abnormal (i.e., faulty) system conditions.

In this paper, an approach to model power system components is introduced which is capable of simulating electromagnetic transients using terminal circuit. The construction of a 5 bus test case is provided to verify the proposed approach.
Adaptive time-stepping method is used to improve computation efficiency of EMT simulation. Balanced three-phase fault is applied to the system for further analysis and the simulation results are compared with commercial software to 
verify that the test case built with terminal circuit is able to simulate electromagnetic transients accurately and efficiently. 

The remainder of the paper is organized as follows: In Section \uppercase\expandafter{\romannumeral2}, terminal circuits are introduced for power system electromagnetic modeling.   
Section \uppercase\expandafter{\romannumeral3} provides details of modeling a 5-bus system, and Section \uppercase\expandafter{\romannumeral4} presents the conclusions. 

\section{Electromagnetic Modelling}
Typically, power system is consisted of generators, trans- mission lines, transformers, and series RLC component, non- ideal voltage source in special cases. They can be further represented by basic circuit elements of resistors, reactors, capacitors, voltage source, current source, and their combination in very large-scale integrated circuit. In this section, power system  component electromagnetic models are constructed by using terminal circuit in VLSI circuit simulator.
\subsection{Park’s Transformation }
\begin{figure}[thpb]
\centerline{\includegraphics[scale=0.45]{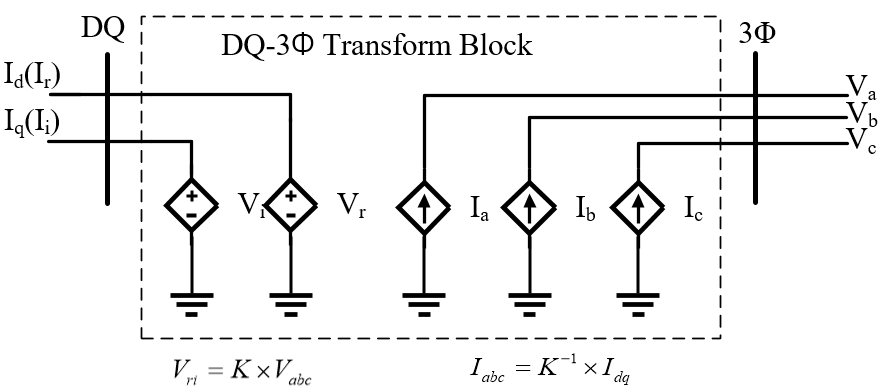}}
\caption{Park's Transformation}
\label{fig:K_transform}
\end{figure}

In EMT studies, generator is modelled by an equivalent current source on the fictitious direct (d) and quadrature (q) axes \cite{kundur1994power} and interfacing with the three-phase network model by Park’s transformation. The Park’s transformation and its reverse transformation are given by (\ref{Park}) (\ref{equ:park_1}), and (\ref{equ:park_2}). And the transform block of Park’s Transformation is shown in Fig. \ref{fig:K_transform}.

\vspace{-2.5mm}
\begin{align} \label{Park}
K &= \sqrt{\frac{2}{3}}
\begin{bmatrix}
\cos(\theta) & \cos(\theta - \frac{2\pi}{3}) & \cos(\theta + \frac{2\pi}{3})\\
\sin(\theta) & \sin(\theta - \frac{2\pi}{3}) & \sin(\theta + \frac{2\pi}{3})\\
\frac{1}{\sqrt{2}} & \frac{1}{\sqrt{2}} & \frac{1}{\sqrt{2}} 
\end{bmatrix}
\end{align}
\vspace{-1.5mm}
\begin{equation} \label{equ:park_1}
V_{ri} = K\times V_{abc}
\end{equation}
\vspace{-5.5mm}
\begin{equation} \label{equ:park_2}
I_{abc} = K^{-1}\times I_{dq} \textcolor{white}{11.1}
\end{equation}

\subsection{Power System Components Modeling}
\begin{figure}[h!]
\centerline{\includegraphics[scale=0.35]{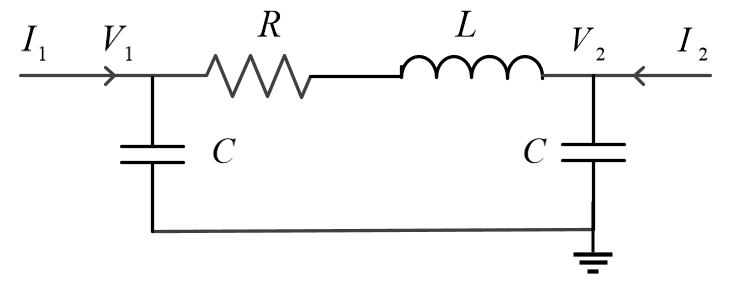}}
\caption{Transmission Line RLC Model}
\label{fig:pi_line}
\end{figure}

In power system, the $\pi$-type transmission line is a typical three phase RLC circuit. The single phase transmission line representation is shown in Fig. \ref{fig:pi_line}.
\vspace{-3.5mm}

\begin{figure}[h]
\centerline{\includegraphics[scale=0.60]{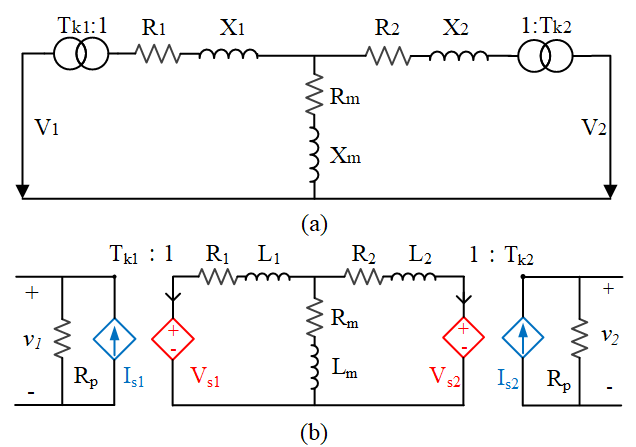}}
\caption{Two-winding Transformer}
\label{fig:Transformer_ratio}
\end{figure}

\begin{figure}[h!]
\centerline{\includegraphics[scale=0.60]{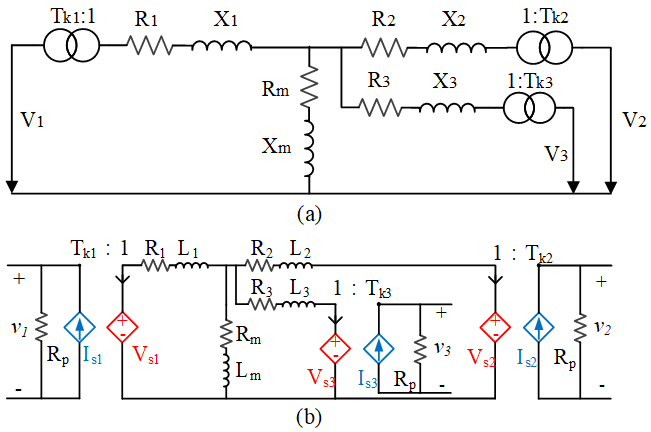}}
\caption{Three-Winding Transformer}
\label{fig:Transformer_3_winding}
\end{figure}

\begin{figure}[h!]
\centerline{\includegraphics[scale=0.54]{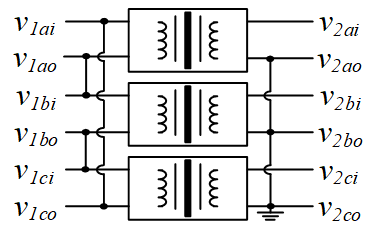}}
\caption{Delta-wye Transformer Connection Diagram}
\label{fig:Delta-wye connection}
\end{figure}

For Transformer, Fig. \ref{fig:Transformer_ratio} (a) shows the equivalent circuit for a single-phase two-winding transformer and the turns ratio of winding 1 (\textit{$T_{k1}$}) and winding 2 (\textit{$T_{k2}$}) is given by (\ref{turns ratio}).
\begin{equation} \label{turns ratio}
T_{k1} (T_{k2}) = \frac{V_{Rated}}{V_{Base}}
\end{equation}
where $V_{Rated}$ is the rated line voltage of winding 1 (2) and $V_{Base}$ is the voltage base of winding 1 (2). In this work, the authors utilize two sets of voltage controlled voltage source to implement the turns ratio  function and two sets of current controlled current source to enforce the current through the mutual impedance, as shown in Fig. \ref{fig:Transformer_ratio} (b). Similarly, the equivalent circuit for a  single-phase three-winding transformer is depicted in Fig. \ref{fig:Transformer_3_winding} (a) and three sets of voltage (current) controlled voltage (current) source are used in this work to implement the voltage and current relationship, as in Fig. \ref{fig:Transformer_3_winding} (b). In addition, a three-phase delta-wye transformer connection is constructed using three sets of single-phase transformers as shown in Fig. \ref{fig:Delta-wye connection}. In this three-phase formulation, the left-hand-side terminals of the transformer show a delta connection with a positive sequence of a-b-c phase while the right-hand-side terminals have a wye connection with all return terminals \textit{$v_{2ao}$}, \textit{$v_{2bo}$}, and \textit{$v_{2co}$} attached to the ground. 

\begin{figure}[h!]
\centerline{\includegraphics[scale=0.5]{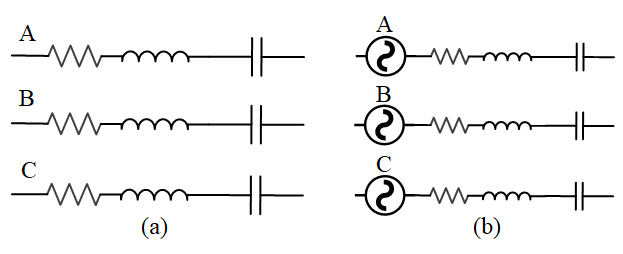}}
\caption{(a) Series RLC (b) Non-Ideal Voltage Source}
\label{fig:RLC}
\end{figure}

In typical power system electromagnetic simulation, three-phase series RLC and the non-ideal voltage source are commonly used, their models are shown in  Fig.\ref{fig:RLC} (a) and (b).

\begin{figure}[h!]
\centering
\includegraphics[scale=0.458]{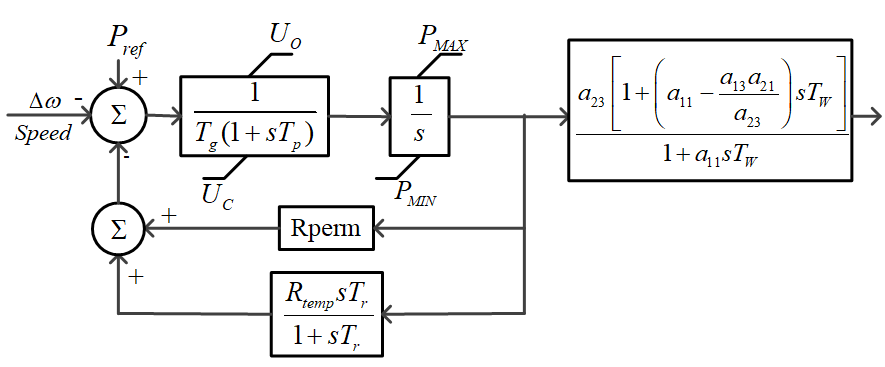}
\caption{IEEEG3 Governor}
\label{fig:IEEEG3}
\end{figure}

\begin{figure}[h!]
\centering
\includegraphics[scale=0.484]{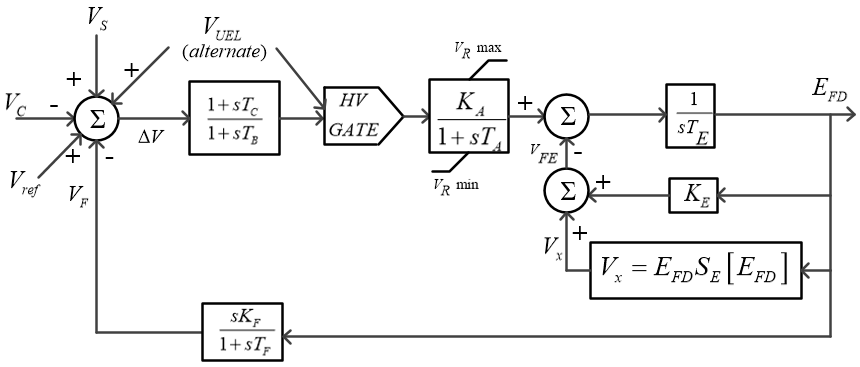}
\caption{DC1A type exciter}
\label{fig:DC1A}
\end{figure}

In electromagnetic transient simulation, 6$^{th}$ order generator model is applied. The associated governor model and exciter model are illustrated as follow. The example transfer function diagrams of a governor and an exciter are shown in  Fig. \ref{fig:IEEEG3} and Fig. \ref{fig:DC1A}, respectively.
 

\begin{figure}[h!]
\centering
\includegraphics[scale=0.595]{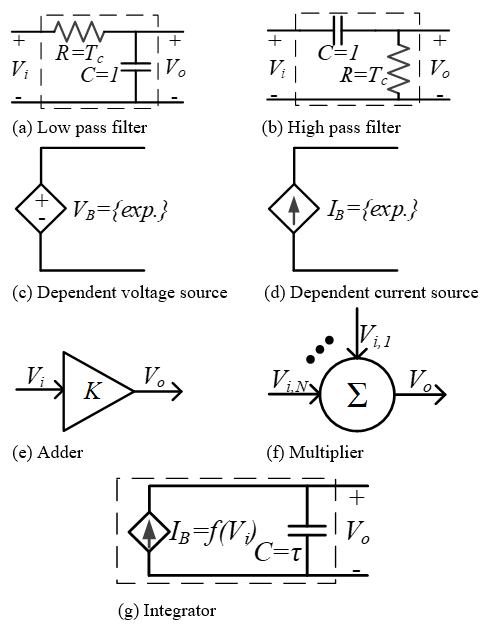}
\caption{Basic Control Block Circuit Model}
\label{fig:Control Blocks}
\end{figure}

The transfer function diagrams are modeled by integrating basic control blocks, such as low pass filter, high pass filter, gain block, adder, integrator etc. Those basic control blocks are modeled by circuit using integrated circuit simulator as shown in Fig. \ref{fig:Control Blocks}.
\vspace{-1.5mm}

While dependent voltage and current source, as shown in Fig. \ref{fig:Control Blocks} (c) and Fig. \ref{fig:Control Blocks} (d), are also known as B-source devices and commonly used as analog behavioral modeling (ABM) in integrated circuit \cite{Xyce}. They are adopted to accommodate complicated nonlinear relations between the input and output signal such as in the integrator model shown in Fig.  \ref{fig:Control Blocks} (g). The integrator circuit is a nonlinear dependent current source in parallel with an ideal capacitor.

\begin{equation}
\frac{dV_0}{dt} = \frac{1}{C}f(V_i)
\end{equation}
where, the nonlinear dependent current source provides a time dependent current $f(V_i)$ that simulates the time derivative of a state variable labeled as $V_o$. The capacitance of the capacitor is set as $\tau$ to make the state variable $V_o$ representing the desired integration.

The lead-lag filter in the IEEEG3 governor and the DC1A type exciter can be modeled by adding the signals of a low pass filter and a high pass filter. And the complicated saturation function $S_E(E_{FD})$ can be modeled by a nonlinear dependent voltage source.

\section{Case Study}
To verify the effectiveness of the proposed method, a simple 5-bus system is modeled by SPICE-level analog circuit and simulated by using an integrated circuit simulator, Xyce \cite{Xyce}.

\begin{figure}[thpb]
\centering
\includegraphics[scale=0.49]{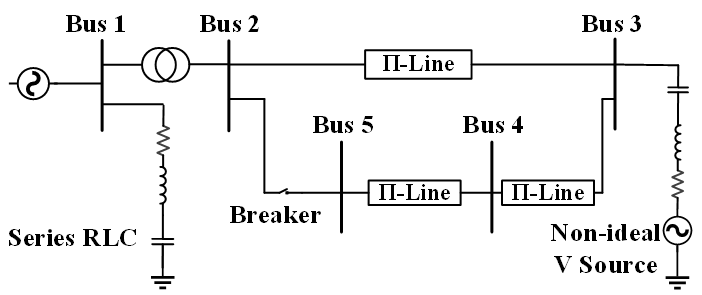}
\caption{5 bus case}
\label{fig:5buscase}
\end{figure}

As shown in Fig. \ref{fig:5buscase}, the 5-bus system contains 1 synchronous generator, 1 non-ideal voltage source, 1 series RLC, 3 $\pi$-type transmission lines and 1 circuit breaker. The fundamental frequency of the system is 50-Hz.
\vspace{-1.5mm}
\vspace{-4.5mm}
\begin{figure}[h]
\centering
\includegraphics[scale=0.59]{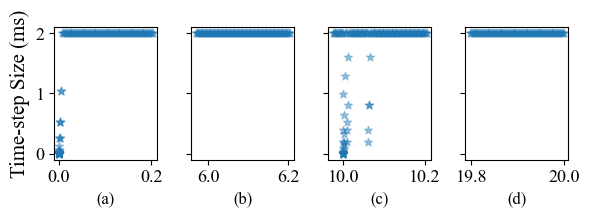}
\caption{Adaptive time-step size according to different activities: small time-step during the initialization $\&$ fault period}
\label{fig:TimeStep}
\end{figure}

To illustrate the electromagnetic transient behavior of the studied system, a balanced three-phase fault is applied at bus 4 at 10.0s and is cleared at \textcolor{black}{10.06s}. In the transient simulation, adaptive time step is applied with a small initial time-step value as 20us and the maximal time-step as 2ms to achieve better simulation efficiency. When the simulation starts, the step size is adapted to increase till the fault happened as shown in   Fig.\ref{fig:TimeStep} (a) and (b). Upon the time of the fault start and clear, the step sizes are reduced to keep acceptable truncation errors during the large transient as shown in  Fig.\ref{fig:TimeStep} (c). After the transient is settled down within seconds after a fault is cleared, as shown in Fig.\ref{fig:TimeStep} (d), the step size increases to the maximal time-step to achieve high computation efficiency. The adaptive time-step balances the computation accuracy and efficiency. By using the adaptive time-step simulation, it takes 2.66s to simulate 20 seconds transients. In contrast, by using fixed time-step at 1ms, it takes 3.98s to complete the 20 seconds transient simulation. 

\begin{figure*}[ht!]
\centering
\includegraphics[scale=0.680]{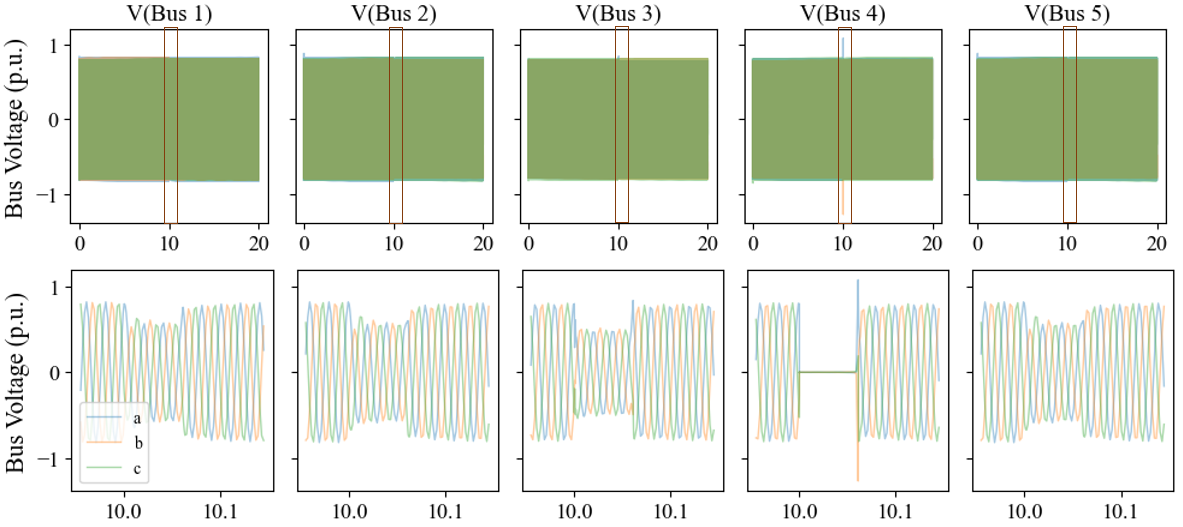}
\caption{Simulation results from the 5 bus case. The fault is applied at 10.00 s and cleared 10.06 s.}
\label{fig:5busVmag}
\end{figure*}

\begin{figure*}[bh!]
\centering
\includegraphics[scale=0.680]{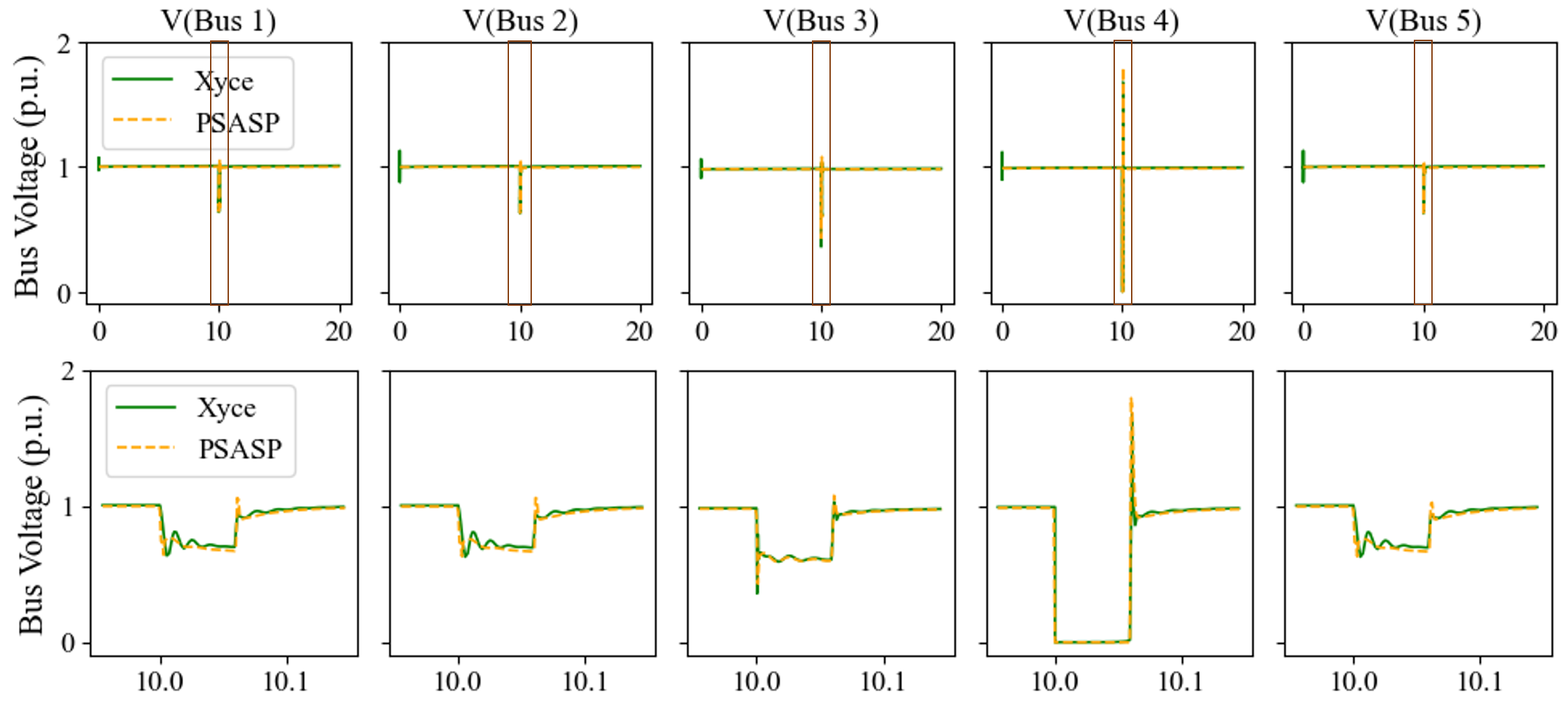}
\caption{Positive sequence voltages of 5 buses collected from the developed case using the Xyce circuit simulator and ADPSS}
\label{fig:5busVmag_compare}
\end{figure*}

To verify the computation accuracy, the simulation results provided in Fig. \ref{fig:5busVmag} are compared with the results simulated by a commercial electromagnetic simulation software Advanced Digital Power System Simulator (ADPSS) \cite{zhongxi1998power} developed by China Electric Power Research Institute (CEPRI) as shown in Fig. \ref{fig:5busVmag_compare}.

As shown in Fig. \ref{fig:5busVmag}, after the fault is applied at 10.0s, the three phase voltages decrease while  the short-circuit current at the faulted bus is increased. Upon the time of the fault removal after three cycles, the voltage spikes caused by phase jump at the clearing time are observed. The fast transients are eliminated within 1 cycle and then rapidly settled down to an equilibrium point. Note that the post-fault three-phase voltage amplitudes are approaching to their pre-fault values.

In Fig. \ref{fig:5busVmag_compare}, the positive sequence voltages at each bus simulated by the proposed method are compared with those collected from ADPSS which demonstrates that the simulation results by the two simulators are consistent.

\section{Conclusion}
In this paper, a new approach is proposed to model power system electromagnetic transients by integrated circuit using SPICE-level analog circuit language. Preliminary studies are carried out to give insights on modeling power system components using terminal circuits.  A Park’s transformation circuit model is developed as an interface between the three-phase network reference frame and the single-phase dq reference frame. A 5-bus case is used as an example to model power system electromagnetic transients by the proposed method. A balanced three-phase fault is applied for the case study. To improve the computation efficiency, adaptive time-step method is used in the simulation. The accuracy of the simulation results are validated by comparing with commercial software. 

This paper proved the concept of using integrated circuit to model power system for the purpose of transient simulation. As enabling technology, the approach and the simulation results verified the possibility and feasibility of using a large-scale integrated circuit chip to model power system and eventually replace the time consuming traditional numerical methods for transient simulation. As demonstrated in the initial practice, the method has a great potential to improve the large-scale power system transient study accuracy and efficiency.



\vspace{12pt}

{
\small
\bibliography{citation}
\bibliographystyle{IEEEtranN}
}
\end{document}